\documentclass[11pt]{article}
\usepackage{amsmath}
\usepackage{latexsym}
\usepackage{amssymb}
\usepackage{mathrsfs,amsmath,amssymb,amsbsy}
\input BoxedEPS
\SetRokickiEPSFSpecial
\HideDisplacementBoxes

\numberwithin{equation}{section}
\title{Diquarks as Inspiration and as Objects}
\author{Frank Wilczek\footnote{Solicited contribution to the Ian Kogan memorial volume, ed. M. Shifman.}}
\begin{document}
\maketitle

\begin{abstract}
Attraction between quarks is a fundamental aspect of QCD.   It is plausible that several of the 
most profound aspects of low-energy QCD dynamics are connected to
diquark correlations, including: paucity of exotics (which is the foundation of the quark model and of traditional nuclear physics),
similarity of mesons and baryons, color superconductivity at high density, hyperfine splittings, 
$\Delta I = 1/2$ rule, and some striking features of structure and fragmentation functions.   After a brief overview of these issues, I discuss how diquarks
can be studied in isolation, both phenomenologically and numerically, and present approximate mass differences for diquarks with different
quantum numbers.   The mass-loaded generalization of the Chew-Frautschi formula provides an essential tool.  
\end{abstract}

\section{Diquarks as Inspiration}

\subsection{Diquarks in Microscopic QCD}

In electrodynamics the basic interaction between like-charged particles is repulsive.   In QCD, however, the primary interaction between two quarks can be attractive.  At the most heuristic level, this
comes about as follows.  Each quark is in the $\bf 3$ representation, so that the two-quark color state $\bf 3 \otimes \bf 3$ can be either the symmetric $\bf 6$ or the antisymmetric 
$\bar {\bf 3}$.    Antisymmetry, of course, is not possible with just 1 color!    Two widely separated quarks each generate the color flux associated with the fundamental representation; if they are
brought together in the $\bf {\bar 3}$, they will generate the flux associated with a single anti-fundamental, which is just half as much.   Thus by bringing the quarks together we lower the gluon
field energy: there is attraction in the $\bf {\bar 3}$ channel.   We might expect this attraction to be roughly half as powerful as the quark-antiquark $\bf 3 \otimes \bar 3 \rightarrow 1$.  Since
quark-antiquark attraction drives the energy in the attractive channel below zero, 
triggering condensation $\langle \bar q q \rangle \ne 0 $ of $q \bar q $ pairs and chiral symmetry breaking, an attraction even half as powerful 
would appear to be potentially quite important for understanding low-energy QCD dynamics.  

One can calculate the quark-quark interaction due to single gluon exchange, and of course one does find that the color $\bf {\bar 3}$ channel for quarks is attractive.  Going a step further, one can
consider magnetic forces, and distinguish the favored spin configuration.  One finds that the favorable spin configuration is likewise the antisymmetric one, i.e. 
$\frac{1}{2}\otimes\frac{1}{2}\rightarrow 0$.    With antisymmetry in color and spin, and a common spatial configuration, Fermi statistics requires that the favorable diquark configuration
is also antisymmetric in flavor.   For non-strange diquarks, this means {\it isosinglet}, in the context of flavor $SU(3)$ it means
flavor $\bar {\bf 3}$.   We shall denote the favorable diquark configuration as $[q q^{\prime}]$, and speak of
``good'' diquark.   We shall also have occasion to consider the spin triplet flavor symmetric configuration (still color  $\bar {\bf 3}$!), which we will denote this as $(q q^{\prime})$ and 
speak of the ``bad'' diquark.   Since the spin-spin interaction is a relativistic effect, we might expect it to be strongest for the lightest quarks; that is, we expect the splitting 
$(ud)-[ud] > (us) - [us] > (uc) - [uc] \approx 0$.

One can also calculate forces between quarks due to instantons.  The same channel emerges as the most favorable, with attraction.

At asymptotically high densities in QCD one can justify the use of weak coupling to analyze quark interactions near the Fermi surface.  
The attractive quark-quark interaction in the good diquark channel is responsible for color superconductivity, and more particularly color-flavor locking. 
In that context it triggers condensation of diquarks, with color symmetry breaking.  This leads to a rich theory, including calculable -- weak coupling, but nonperturbative -- 
mechanisms for confinement and chiral symmetry breaking.  In vacuum we do not have color breaking, of course, or (therefore) diquark condensation; 
but the dominant role of good diquarks at 
high density is definitely another motivation for studying their properties in general.   {}As a practical matter, it might help us understand the parameters governing the approach to
asymptopia, which is important for constructing models of the internal structure of neutron ($\rightarrow_{{\rm center}}$ quark) stars.  

As a corollary to the fact that quark attraction that favors good diquark formation, we might expect {\it repulsion between good diquarks}.   Indeed, when two good diquarks overlap 
the cross-channels, involving one quark from each diquark, will have unfavorable correlations.   The repulsion might be manifested in the form of a force or, in response to
attempts at fusion, re-arrangement into baryon plus single quark.

\subsection{Phenomenological Indications}

These heuristic, perturbative, and quasi-perturbative considerations suggest several ``applications'' of diquark ideas within strong interaction phenomenology.   Since the relevant calculations are
not performed in a well-controlled approximation, we should regard this as an exploratory activity.   To the extent that we discover interesting things in this way -- and we do! -- it poses the challenge 
of making firmer, more quantitative connections to fundamental theory.

A classic manifestation of energetics that depends on diquark correlations is the $\Sigma-\Lambda$ mass difference.   The $\Lambda$ is isosinglet, so it features $[ud]$; while $\Sigma$, being
isotriplet, features $(ud)$.   The $\Sigma$ is indeed heavier, by about 80 MeV.  Of course, this comparison of diquarks 
is not ideal, since the spectator $s$ quark also has significant spin-dependent interactions.  A cleaner 
comparison involves the charm analogues, where $\Sigma_{c} - \Lambda_{c} = 215$ MeV.   (Actually this comparison is not so clean either, as we'll discuss later.  One sign of uncleanliness is that
there either $\Sigma_{c}(2520)\frac{3}{2}^{+}$ or $\Sigma_{c}(2455)\frac{1}{2}^{+}$ might be used for comparison; here I've taken the weighted average.)

One of the oldest observations in deep inelastic scattering is that the ratio of neutron to proton structure functions approaches $\frac{1}{4}$ in the limit $x\rightarrow 1$
\begin{equation}\label{structureRatio}
\lim_{x\rightarrow 1} \frac{F_{2}^{n}(x)}{F_{2}^{p}(x)}  \rightarrow \frac{1}{4} 
\end{equation}   
{}In terms of the twist-two operator matrix elements used in the formal analysis of deep inelastic scattering, this translates into the statement
\begin{equation}\label{opValence}
\lim_{n\rightarrow \infty} \frac{\langle p| \bar d \gamma_{\mu_1}\overleftrightarrow\nabla_{\mu_2}\cdots \overleftrightarrow\nabla_{\mu_n} d | p \rangle}
{\langle p| \bar u \gamma_{\mu_1}\overleftrightarrow\nabla_{\mu_2}\cdots \overleftrightarrow\nabla_{\mu_n} u | p \rangle} \rightarrow 0
\end{equation}
where spin averaging of forward matrix elements, symmetrization over the $\mu$s, and removal of traces is implicit, and a common tensorial form is factored out, together
with similar equations where operators with strange quarks, gluons, etc.  appear in the numerator.   
Equation (\ref{opValence}) states
that in the valence regime $x\rightarrow 1$, where the struck parton carries all the longitudinal momentum of the proton, that struck parton must be a $u$ quark.   
It implies, by isospin symmetry, the corresponding 
relation for the neutron, namely that in the valence regime within a neutron the parton must be a $d$ quark.  Then the ratio of neutron to proton matrix elements will be governed by the ratio of the squares
of quark charges, namely $\frac{(-\frac{1}{3})^{2}}{(\frac{2}{3})^{2}} = \frac{1}{4}$.  Any (isosinglet) contamination from other sources will contribute equally to numerator and denominator, 
thereby increasing this ratio.   Equation (\ref{opValence}) is, from the point of view of symmetry, a peculiar relation: it requires an emergent conspiracy between isosinglet and isotriplet operators.   It is, from
a general physical point of view, most remarkable: it is one of the most direct manifestations of the fractional charge on quarks; and it is a sort of hadron = quark identity, closely related to the
quark-hadron continuity conjectured to arise in high density QCD.   It is an interesting challenge to derive (\ref{opValence}) from microscopic QCD, and to estimate the rate of approach to 0.  

A more adventurous application is to fragmentation.  One might guess that the formation of baryons in fragmentation of an energetic quark or gluon jet could proceed stepwise, through the formation of 
diquarks which then fuse with quarks.   To the extent this is a tunneling-type process, analogous to pair creation in an electric field, induced by the decay of color flux tubes, one might expect that the good 
diquark would be significantly more likely to be produced than the bad diquark.   This would reflect itself in a large $\Lambda/\Sigma$ ratio.   And indeed, data from LEP indicates that the value of this ratio is
about 10 at large $z$.  In the Particle Data Book one also finds an encouraging ratio for total multiplicities in $e^{+}e^{-}$ annihilation: $\Lambda_{c}:\Sigma_{c} = .100\pm .03: .014 \pm .007$; in this
case the $c$ quarks are produced by the initiating current, and we have a pure measure of diquarks.

There are also several indications that diquark correlations have other important dynamical implications.  
The $\Delta I = \frac{1}{2}$ rule in strangeness-changing nonleptonic decays has also been ascribed to attraction in the diquark channel.  The basic operator
$\bar u \gamma_{\mu}(1-\gamma_{5})d \bar s \gamma^{\mu} (1-\gamma_{5}) u$ arising 
from $W$ boson exchange can be analyzed into $\bar{[us]}[ud]$, $\bar{(us)}(ud)$, and related 
color-{\bf{6}} diquark types.   Diquark attraction in $\bar{[us]}[ud]$ means that there is a larger chance for quarks in this channel to tap into short-distance components of hadronic wavefunctions.  
This effect is reflected
in enhancement of this component of the basic operator as it is renormalized toward small momenta.  Such an enhancement is well-known to occur at one-loop order (one gluon exchange).   Stech and Neubert
have advanced this line of thought significantly \cite{stech}.

\subsection{Correlations and the Main Problem of Exotics}

Our present understanding of the strong interaction is disturbingly schizophrenic.  
On the one hand we have an algorithmically definite
and very tight relativistic quantum field theory, 
quantum chromodynamics (QCD), which we can use to do accurate quantitative calculations in special circumstances.   Many hard 
(i.e., large momentum-transfer) processes and processes involving heavy quarks can be treated using the techniques of perturbative QCD.   The 
spectroscopy of low-lying states, and a few interesting matrix elements of operators (currents, twist-two operators, weak Hamiltonian matrix elements) 
can be calculated by direct numerical solution of the fundamental equations, using the techniques of lattice gauge theory.  These quantitative
calculations are famously successful, with accuracies approaching 1\% in favorable cases, and amply justify faith in the theory.   The basic degrees of
freedom in QCD include massless gluons and almost-massless $u,d$ quarks, and the interaction strength, though it ``runs'' to small coupling at large 
momentum transfer, is not uniformly small.   We might therefore anticipate, heuristically, 
that low-energy gluons and quark-antiquark pairs are omnipresent, and in particular that the eigenstates of the Hamiltonian -- hadrons -- will be complicated 
composites, containing an indefinite number of particles.   And indeed, according to the strictest experimental measure of internal structure available, the structure 
functions of deep inelastic scattering, nucleons do contain an infinite number of soft gluons and quark-antiquark pairs (parton distributions $\sim \frac{dx}{x}$ as $x\rightarrow 0$).

The quark model has been used with considerable success to organize a lush jungle of observations that would otherwise appear bewildering.   It is built upon degrees of freedom whose
properties are closely modeled on those of the fundamental theory; nevertheless, its success raises challenging conceptual questions.   
{}For the main working assumption of the quark model is that
hadrons are constructed according to two body plans: mesons, consisting of a quark and an antiquark; and baryons, consisting of three quarks.  This seems out of step with the heuristic expectations we
mentioned earlier.   

And, lest we forget, the most developed and useful model in strong interaction physics is traditional nuclear physics, based on nucleons as degrees of freedom.  In this model 
the effective 
residual interactions are feeble compared to the interactions responsible for constructing the nucleons from massless ingredients in the first place; this allows us to employ essentially non-relativistic dynamics,
and we don't consider particle production.    Furthermore, and not unrelated: the nuclear forces have a ``hard core'' repulsion, and saturate.  

The puzzles posed by the success of the quark model and traditional nuclear physics are sharply posed in the question of {\it exotics}.   Are there additional body plans in the hadron spectrum, 
beyond $qqq$ baryons and $\bar q q$ mesons (and loose composites thereof)?  If not, why not; if so, where are they?   As a special case: why don't multi-nucleons merge into single bags, e.g. 
$qqqqqq$ -- or can they?

The tension between {\it a priori\/} expectations of complex bound states and successful use of simple models, defines the main problem of 
exotics: Why aren't there more of them?  A heuristic explanation can begin along the following 
lines.   Low-energy
quark-antiquark pairs are indeed abundant inside hadrons, as are low-energy gluons, but they have (almost) vacuum quantum numbers: they are arranged in flavor and spin singlets.  
(The ``almost'' refers to chiral symmetry breaking.)   Deviations from the ``good'' quark-antiquark or gluon-gluon channels, which are color and spin singlets, cost significant energy.  States which contain such
excitations, above the minimum consistent with their quantum numbers,  will tend to be highly unstable.    They might be hard to observe as resonances, or become unbound altogether.

The next-best way for extraneous quarks to organize themselves appears, according to the preceding considerations, to be in ``good'' diquark pairs.   Thus a threatening -- or promising -- strategy for constructing
low-energy exotics apparently could be based on using these objects as building-blocks.   There are two reasons  ``good'' diquark correlations help explain the paucity of exotics: because of their 
antisymmetry, they lock up spin and
flavor; and because of their repulsion, they forbid mergers.   These two aspects are exemplied in the next two paragraphs.

Tetraquarks play an important role in modeling the observed low-lying nonet of scalar $0^{+}$ mesons including $f_{0}(600)= \sigma, \kappa(900), f_{0}(980), a_{0}(980)$.   It appears perverse to model
these as conventional $q \bar q $ mesons, since the isotriplet $a_{0}(980)$ is the heaviest component, but would (on this assignment) contain no strange quarks.   A serious and extensive case has been made that
an adequate model of these mesons must include a major admixture of $ q q \bar q \bar q$.   Then both $f_{0}(980), a_{0}(980)$ are accommodated as $[ls]\bar {[ls]}$, with $l= u~{\rm or}~d$.
{}For our purposes, the most important observation is that if the quarks (antiquarks) are correlated into 
good diquarks (antidiquarks), as we expect they will be for the lowest-lying states, then the {\it non}-exotic flavor structure of the nonet is explained; indeed, for the flavor one obtains 
$\bar {\bf 3}\otimes {\bf 3} = {\bf 8} \oplus 1$ with the same charges as for $ q \bar q $.   {}For this reason they are called cryptoexotics.   $ q q \bar q \bar q$ can organize alternatively into 
two color singlet $q\bar q$ mesons, of course, and sophisticated modeling includes both channels (with diquarks dominating at short distances, mesons at larger distances).

The non-existence of low-lying dibaryons is related to the (or at least, a) foundational problem of nuclear physics: Why do protons and neutrons in close contact retain their integrity?  
Essentially the same question arises in a sharp form for the $H$ particle studied by Jaffe \cite{jaffe}.   It has the configuration $uuddss$.   
In the bag model it appears that a single bag containing these quarks supports a spin-0 state that is quite favorable energetically.   A calculation based on quasi-free quarks 
residing in a common bag, allowing for 
one-gluon exchange, indicates that $H$ might well be near or even below $\Lambda \Lambda$ threshold, and thus strongly stable; or perhaps even below $\Lambda n$ threshold, and therefore stable even against
lowest-order weak interactions.   These possibilities appear to be ruled out both experimentally and by numerical solution of QCD, though possibly neither case is airtight.    Good diquark correlations, together with
repulsion between diquarks, suggests a reason why the almost-independent-particle approach fails in this case.    Note that for this mechanism to work requires that essentially nonperturbative 
quark interaction effects, beyond one gluon exchange, must be in play.

\section{Diquarks as Objects}

{}From all this it appears that diquarks may be very useful degrees of freedom to focus on in QCD.   If we're going to do that, the first step should be to study them in a pure and
isolated form, and determine their parameters.    This is not straightforward, due to confinement, since the diquarks are colored.   
But I believe there are attractive ways to do something approaching isolating them, both physically and
numerically.   

Of course, the same problem arises for quarks.  Our considerations will apply to them in a non-trivial way, as well.

In rapidly spinning baryons centrifugal forces
lead to a geometry where a quark at one end of a line of color flux is joined to two quarks at the other.   The two-quark end then makes a little laboratory where one can
compare good and bad diquark configurations with each other, assess the effects of strangeness,  and (comparing with mesons) normalize them relative to single quarks.  

Famously, the Chew-Frautschi formula
\begin{equation}
M^{2} = a + \sigma L
\end{equation}
organizes trajectories of resonances (Chew-Frautschi formula) with the same internal quantum numbers but different values of 
$J^{P}$; here $\sigma$ is a universal constant $\sim 1.1$ Gev$^{2}$ while $a$ depends on the quantum numbers, and  $L$
is an orbital angular momentum, quantized in integers.   Recently Alex Selem and I have used this formula, together with some refinements and extensions, to do extensive and I think 
quite successful hadron systematics.  
My main point below, extracted from that work, will take off from one such refinement.   

The formula $M^{2}= \sigma L$ arises from solving the equations for a spinning relativistic string with tension $\sigma/(2\pi)$, terminated by the
boundary condition that both ends move transversely at the speed of light.  We might expect it to hold asymptotically for large $L$ in QCD, when an elongated
flux tube appears string-like, the rotation is rapid, quark masses are negligible, and semiclassical quantization of its rotation becomes appropriate.    
The primeval CF formula $M^{2} = a +  \sigma L$, with simple non-zero values of $a$
(e.g., $a=\frac{1}{2}\sigma$) can result from quantization of an
elementary non-interacting string, including zero-point energy for string vibrations. 

In the following section we (that is, Alex and I) generalize 
the classical formula to the form appropriate for string termination on massive objects.   There will be corrections that depend on the masses of the objects at the end.   Using these corrected formulas,
we are able to identify (over-determined) values of the masses of various kinds of quasi-isolated quarks and diquarks, directly from spectroscopic data.

\subsection{Generalization of the Chew-Frautschi Formula}

We can generalize the Chew-Frautschi formula by considering two masses $m_1$, $m_2$ connected by a relativistic string with constant tension, $T$, rotating with angular momentum $L$. 
Our general solution naturally arises in a parameterized form in which the energy,$E$, and $L$ are both expressed in terms of the angular velocity, $\omega$, of the rotating system. 
In the limit that $m_1$, $m_2$ $\rightarrow 0$, the usual Chew-Frautschi relationship $E^2 \propto L$ appears.

Considering masses $m_1$ and $m_2$ at distances $r_1$ and $r_2$ away from the center of rotation respectively.  The whole system spins with angular velocity $\omega$. 
It is also useful to define:
\begin{equation}
\gamma_i= \frac{1}{\sqrt{1-(\omega r_i)^2}}
\label{eq:gam}
\end{equation}  
where the subscript $i$ can be 1 or 2 (for the mentioned masses). It is straightforward to write the energy of the system:
\begin{equation}
E= m_1\gamma_1 +  m_2\gamma_2 + \frac{T}{\omega}\int_0^{\omega r_1}\frac{1}{\sqrt{1-u^2}}du + \frac{T}{\omega}\int_0^{\omega r_2}\frac{1}{\sqrt{1-u^2}}du.
\label{eq:energy1}
\end{equation}
The last two terms are associated with the energy of the string. Similarly, the angular momentum can be written as: 
\begin{equation}
L= m_1\omega r_1^2\gamma_1 + m_2\omega r_2^2\gamma_2 + \frac{T}{\omega^2}\int_0^{\omega r_1}\frac{u^2}{\sqrt{1-u^2}}du + 
\frac{T}{\omega^2}\int_0^{\omega r_2}\frac{u^2}{\sqrt{1-u^2}}du.
\label{eq:angmom1}
\end{equation}
Carrying out the integrals gives:
\begin{subequations}
\label{eq:EandL}
\begin{eqnarray}
E &=& m_1\gamma_1 + m_2\gamma_2 + \frac{T}{\omega}(\arcsin[\omega r_1] + \arcsin[\omega r_2]), \\
L& =& m_1\omega r_1^2\gamma_1 + m_2\omega r_2^2\gamma_2 \\
 && + \frac{T}{\omega^2}\frac{1}{2} \left( -\omega r_1\sqrt{1-(\omega r_1)^2} + \arcsin[\omega r_1] - \omega r_2\sqrt{1-(\omega r_2)^2} +  \arcsin[\omega r_2]\right) \nonumber 
 \,.
\end{eqnarray}
\end{subequations}
Furthermore the following relationship between the tension and angular acceleration holds for each mass:
\begin{equation}
m_i\omega^2 r_i=\frac{T}{\gamma_i^2}.
\label{eq:forces}
\end{equation}
We can use this to eliminate the distances $r_1$ and $r_2$ and express everything in terms of $\omega$. Specifically we note that in our expressions for $E$ and $L$, 
the quantities that contain $r_i$, are $\gamma_i$ and also $\omega r_i$. From equation~(\ref{eq:forces}) we can ultimately solve for $\gamma_i$:
\begin{equation}
\gamma_i=\sqrt{\frac{1}{2} + \frac{\sqrt{1+ 4(T/(m_i\omega))^2}}{2}}.
\label{eq:gamw}
\end{equation} 
From equation~(\ref{eq:forces}) we also know that $\omega r_i$ is just $T/(m_i\gamma_i^2\omega)$.

We are now in a position to replace these terms in equation~(\ref{eq:EandL}) and write $E$ and $L$ in terms of the parameter $\omega$ and other quantities assumed known, 
namely the masses and the string tension $T$. 
The resulting expressions are a bit opaque, but we can make good use of them either by plotting $E^2$ vs $L$ parametrically, 
or by making appropriate expansions, for the cases of either very light or very heavy masses, to obtain analytic expressions for $E^2$ vs $L$.

The terms associated with each mass decouple from one another, so we may construct expansions for each separately.  
We adopt the convention that the contribution from one mass is preceded with a $\delta$, as in $\delta E$.  It is useful to define another variable $x_i\equiv \frac{m_i\omega}{T}$, 
If we expand in $x_i$, then, we find the contribution to the energy, $\delta E$, and angular momentum, $\delta L$,  due to one light mass is 
\begin{subequations}
\label{eq:mexpand}
\begin{equation}
\delta E_{light} = \frac{\pi T}{2\omega}+ \frac{1}{3} m_i^{1/2} x_{i}^{1/2} + \frac{1}{20} m_i^{1/2} x_{i}^{3/2} + O\left(m_i^{1/2}x_i^{5/2}\right)
\end{equation}
\begin{equation}
\delta L_{light} = \frac{\pi T}{4\omega^2} -
\frac{1}{3}\frac{m_i}{\omega } x_{i}^{1/2}+
\frac{3}{20}\frac{m_{i}}{\omega} x_{i}^{{3/2}} + O\left(\frac{m_{i}}{\omega} x_{i}^{5/2}\right).
\end{equation}
\end{subequations}
to order $x_i^{3/2}$.  For a system with two light and equal masses, we would of course just multiply the right hand side of these expressions by two to obtain the total energy and angular momentum. 
Note that for a very light mass it appears that  $\omega\rightarrow\infty$ as $L\rightarrow 0$, so this is a singular limit.

If we let both masses go to zero, and therefore take only the first term for each mass from the light-mass expansion (equation~(\ref{eq:mexpand})), 
then we recover the familiar Chew-Frautschi relationship for the string with massless ends:
\begin{equation}
\label{eq:lightlight}
E^2 = (2\pi T)L 
\end{equation}

For the first corrections at small $m_{1}, m_{2}$ (and $L\ne 0$) we find, after some algebra,
\begin{equation}\label{greatFormula1}
E \approx \sqrt {\sigma L} + \kappa L^{-\frac{1}{4}} \mu^{\frac{3}{2}}
\end{equation}
with 
\begin{equation}\label{greatFormula2}
\kappa \equiv \frac{2}{3} \frac{\pi^{\frac{1}{2}}}{\sigma^{\frac{1}{4}}}
\end{equation}
and 
\begin{equation}\label{greatFormula3}
\mu^{\frac {3}{2}} \equiv  m_1^{\frac {3}{2}} + m_2^{\frac {3}{2}}
\end{equation}
This is a useful expression, since it allows us to extract expressions for quark and diquark mass differences from the observed values of baryon and meson mass differences.  
Numerically, $\kappa \approx 1.15$ GeV$^{-\frac{1}{2}}$ for $\sigma \approx 1.1$ GeV$^{2}$.  

For heavy-light systems the corresponding formula is
\begin{equation}
E-M = \sqrt{\frac{\sigma L}{2}} + 2^{\frac{1}{4}} \kappa L^{-\frac{1}{4}} \mu^{\frac{3}{2}}
\end{equation}
where $M$ is the heavy quark mass and $\mu$ is the light quark mass.

Note that the usual correction due to a zero-point vibrations, i.e. a classic intercept of the type $E^{2} = a + (2\pi T)L $, yields corrections of the form 
$E \rightarrow \sqrt {\sigma L} + \frac{a}{2\sqrt {\sigma L}}$.  It becomes subdominant to mass corrections at large $L$.  

\subsection{Nucleon-Delta Complex}

As a small taste of the much more extensive analysis presented in \cite{sw},  our
fit to the bulk of non-strange light baryons is presented in Table 1. 
The entries contain central values of masses as quoted in the Particle Data Tables, together with
spin-parity assignments.   By definition nucleons have isospin $\frac{1}{2}$, deltas have isospin $\frac{3}{2}$.   We have included only resonances rated $2^{*}$ or better.   

\begin{table}
\begin{center}
\begin{tabular}{|c|c|c|}
\multicolumn{3}{c}{\bf I. Maximal spin alignment for ``good'' and ``bad'' diquarks}\\[1mm]
\hline
{\it Angular} &{\bf A.~~[ud]---l}&{\bf B.~~(ud)---l}\\
{\it Momentum (L)} &$-$---$\uparrow$&$\Uparrow$---$\uparrow$\\
\hline
0&$N(939)~1/2^{+}$& $\Delta(1232)~3/2^{+}$~~~~~~~~~~~~~~~~~~~~~~\\
1&$N(1520)~3/2^{-}$&  ~~~~~~~~~~~~~~~~~~~~~~~~$N(1675)~5/2^{-}$\\
2&$N(1680)~5/2^{+}$&~$\Delta(1950)~7/2^{+}~~~~N(1990)~7/2^{+}$\\
3&&~~$\Delta(2400)~9/2^{-}~~~~N(2250)~11/2^{-}$\\
4&$N(2220)~9/2^{+}$& $\Delta(2420)~11/2^{+}$~~~~~~~~~~~~~~~~~~~~~~\\
5&$N(2600)~11/2^{-}$& $\Delta(2750)~13/2^{-}$~~~~~~~~~~~~~~~~~~~~~~\\
6&$N(2700)~13/2^{+}$& $\Delta(2950)~15/2^{+}$~~~~~~~~~~~~~~~~~~~~~~\\
\hline
\end{tabular}
\vskip 3mm
\begin{tabular}{|c|c|c|}
\multicolumn{3}{c}{\bf II. ``Bad'' diquark with net spin 1 anti-aligned and}\\
\multicolumn{3}{c}{\bf  ~~~~~``good'' diquark with net spin 1 anti-aligned}\\[1mm]
\hline
{\it Angular} &{\bf A.~~[ud]---l}&{\bf B.~~(ud)---l}\\
{\it Momentum (L)} &$-$---$\downarrow$&$\Uparrow$---$\downarrow$ or $\Leftrightarrow$---$\uparrow$\\
\hline
1&$N(1535)~1/2^{-}$&~$\Delta(1700)~3/2^{-}~~~~N(1700)~3/2^{-}$\\
2&$N(1720)~3/2^{+}$&~$\Delta(1905)~5/2^{+}~~~~N(2000)~5/2^{+}$\\
&&~$\Delta(2000)~5/2^{+}$~~~~~~~~~~~~~~~~~~~~~~~\\
3&&  ~~~~~~~~~~~~~~~~~~~~~~~~$N(2190)~7/2^{+}$\\
4&& $\Delta(2300)~9/2^{+}$~~~~~~~~~~~~~~~~~~~~~~\\
\hline
\end{tabular}
\vskip 3mm
\begin{tabular}{|c|c|}
\multicolumn{2}{c}{\bf III. ``Bad'' diquark with net spin 2 anti-aligned}\\[1MM]
\hline
{\it Angular} &{\bf A.~(ud)---l}\\
{\it Momentum (L)} &$\Downarrow$---$\uparrow$ or $\Leftrightarrow$---$\downarrow$\\
\hline
1&~$\Delta(1620)~1/2^{-}~~~~N(1650)~1/2^{-}$\\
2&~$\Delta(1920)~3/2^{+}~~~~N(1900)~3/2^{+}$\\
3&~~~~~~~~~~~~~~~~~~~~~$N(2200)~5/2^{-}$\\
\hline
\end{tabular}
\vskip 3mm
\begin{tabular}{|c|c|}
\multicolumn{2}{c}{\bf IV. ``Bad'' diquark with net spin 3 anti-aligned}\\[1MM]
\hline
{\it Angular} &{\bf A.~~(ud)---l}\\
{\it Momentum (L)} &$\Downarrow$---$\downarrow$ \\
\hline
2&$\Delta(1910)~1/2^{+}~~~~~~~~~~~~~~~~~~~~~~$\\
3&~~~~~~~~~~~~~~~~~~~~~~~~$N(2080)~3/2^{-}$\\
\hline
\end{tabular}
\caption{Fit to nucleon and delta resonances, based on the standard baryon body plan}
\end{center}
\end{table}
\vspace{6mm}

 The first series assumes maximal alignment between orbital and spin angular momentum.  For $L=0$ there is a unique nucleon state, since (assuming spatial symmetry) spin symmetry and color antisymmetry 
imply flavor symmetry.  For larger values of $L$ there is both a good diquark and a bad diquark nucleon state.  The latter is made by assembling the $I =1$ bad diquark with the $I=\frac{1}{2}$ 
quark to make
$1\otimes \frac{1}{2} \rightarrow \frac{1}{2} $.   Anticipating dynamical independence of the two ends, we should expect to have approximately degenerate bad diquark nucleons and
deltas.  There are many examples of this phenomenon, as we shall see shortly, but only two appear in the first series (and one of those is corrupted).   
The existence of a second nucleon series is a profound fact: it means
that there really is a ``2 against 1'' structure for the quarks, as opposed to a common spatial wave function for all 3, which we encountered for $L=0$.  
In the language of chemistry, we might say it is evidence for a valence-bond, as opposed to a molecular orbital, 
organization.  

A clear distinction between the masses of good versus bad diquark states is visible, upon comparing the first column to the second and third.  The splitting between these states is about 200 MeV.  

There are gaps in the table for a spin-parity $\frac{5}{2}^{-}$ delta around 1700 MeV, a spin-parity $\frac{7}{2}^{+}$ nucleon around 2000 MeV, and possibly for high-spin nucleons to continue the
third column.   The $\Delta(2400)$ would be more comfortable if it were lighter by $\sim$ 100 MeV.   These may be taken as predictions.   

In fitting the good nucleon series even roughly to a formula of the CF form $M^{2} = a + \sigma L$ we discover that it is necessary to separate even and odd $L$.  
We will discuss a
possible microphysical origin for this separation momentarily below, in a separate subsection.

We will give less textual detail in describing the remaining series, since most of the required explanation is so similar.  

 \begin{figure}
\BoxedEPSF{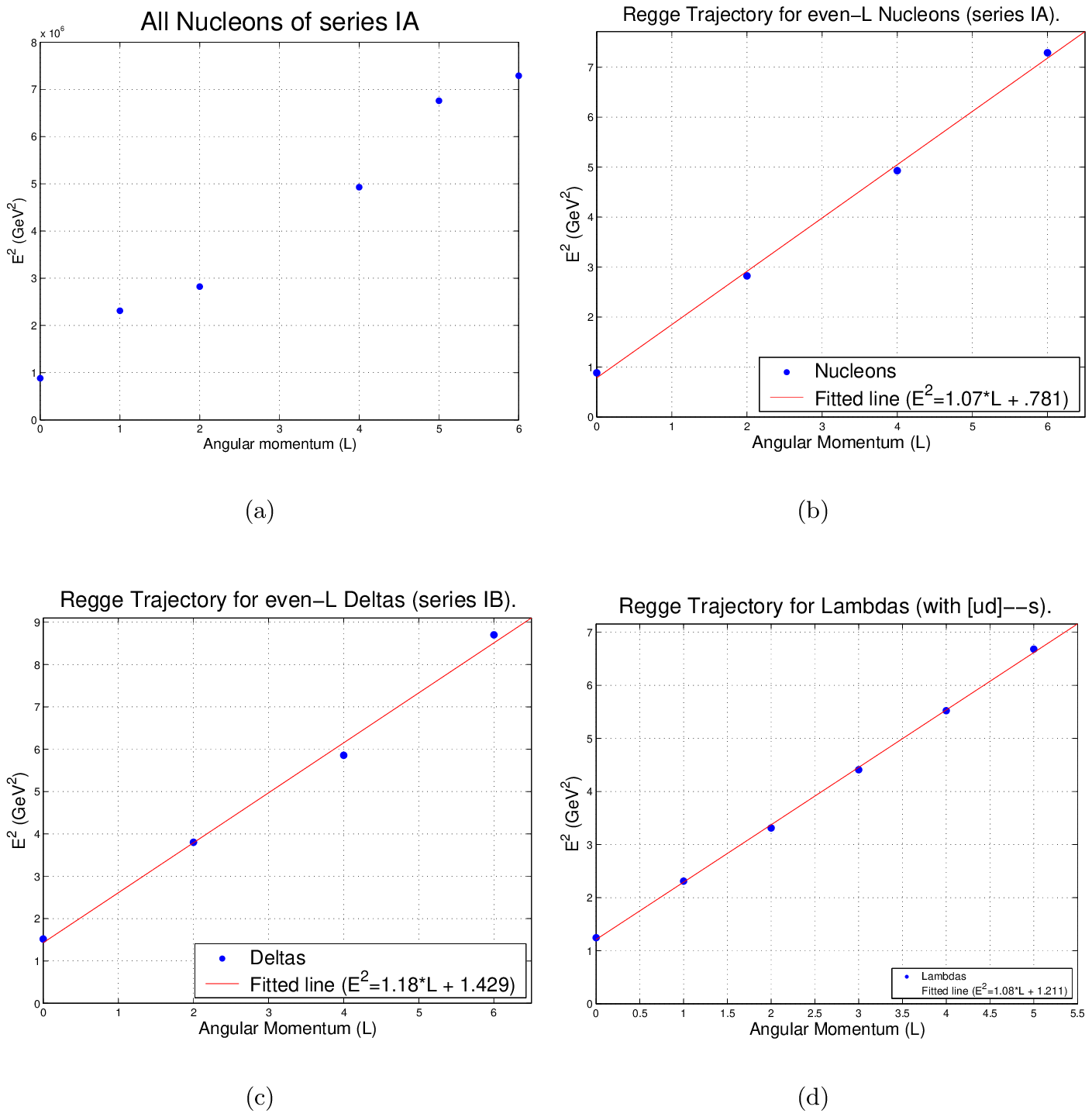}
\caption{Various $E^2$ vs $L$ plots. (a) is a plot of all nucleons of series IA, showing ``even-odd effect". (b-d) are plots of prominent Regge trajectories.}
 \label{fig:regge}
\end{figure}

The second series includes cases where the spin and orbital angular momenta sum up to one less than the maximum possible $J$.  It starts at $L=1$.  
There is a unique good diquark nucleon series, corresponding
to the second term in
\begin{equation}\label{spinCouplingGood} 
L\otimes \frac{1}{2} =  (L + \frac{1}{2}) \oplus (L - \frac{1}{2})
\end{equation}  
but two bad diquark series, corresponding to the second and third terms in 
\begin{equation}\label{spinCouplingsBad}
L\otimes 1 \otimes \frac{1}{2} = (L+\frac{3}{2}) \oplus (L+\frac{1}{2}) \oplus (L + \frac{1}{2}) \oplus (L - \frac{1}{2}) \oplus (L- \frac{1}{2}) \oplus (L-\frac{3}{2})
\end{equation}
(with of course the understanding that negative values are to be dropped, and that for $L=0$  $L + \frac{1}{2}$ occurs only once).  
For $L=0$ there is no clean separation of two ends, and hence no effective approximate isospin conservation to stabilize the
bad diquark; so the absence of those states is not surprising.  
The only case where a doubling is apparent is for the $\frac{5}{2}^{+}$ $\Delta(1905), \Delta(2000)$; we predict that
there are many additional doublets yet to be resolved.  (In our fit to the meson sector, several doublets of this kind appear.)

Beginning with the third series we should not, and do not, find a good-diquark nucleon column.  

The fourth series is very poorly represented; this is not wholly unexpected, since it is predicted to start at $L=2$.   

The surprising feebleness of spin-orbit forces manifests itself most abundantly for $L=2$.   We find two nearly degenerate good-diquark nucleons 
$N(1680), N(1720)$ with 
$J^{P} = \frac{5}{2}^{+}, \frac{3}{2}^{+}$; 
and a host of nearly degenerate bad-diquark nucleons and deltas: 
$N(1990) \frac{7}{2}^{+}$, $N(2000) \frac{5}{2}^{+}$, $N(1900)\frac{3}{2}^{+}$, 
$\Delta(1950)\frac{5}{2}^{+}$, $\Delta(1905) \frac{5}{2}^{+}$, $\Delta(2000)\frac{5}{2}^{+}$,
$\Delta(1920)\frac{3}{2}^{+}$, $\Delta(1910) \frac{1}{2}^{+}$!    


\subsubsection{Even-Odd Effect and Tunneling}

We have mentioned that the even and odd $L$ members of a sequence representing different rotational states of a bone with given internal quark-structure can lie on different trajectories.
A possible microphysical explanation for this is connected with the possibility of quark tunneling from one bone-end to the other.   Imagining the bone in a fixed position, such tunneling produces the
same effect as rotation through $\pi$.   We should construct internal spatial wave-functions which are symmetric or antisymmetric under this interchange.
The former will be nodeless, and lower in energy than the latter, which have a node.   The symmetric states will allow only even $L$, the antisymmetric states will allow only odd $L$.  Thus if tunneling of
this kind is significant we should expect an even-odd splitting the trajectory, with the odd component elevated.   This is what we observe in the nucleon and delta trajectories.  (For this and
the subsequent related assertions, see Figure (\ref{fig:regge})).
A larger effect might be expected for the trajectories with bad diquarks, since the ends won't be sticky.   This too is what is is observed.  

In the $\Lambda$ trajectory the dominant quark configuration has $[ud]$ on one end and $s$ on the other.  It requires triple tunneling to mimic the effect of a $\pi$ rotation.  Thus
we do not expect an even-odd effect here, and none is evident in the data, which has entries for $L=0$ through 5.

\subsection{Results and Conclusion} 

By comparing good nucleons with the corresponding bad nucleons and deltas, using Equations (\ref{greatFormula1}, \ref{greatFormula2}, \ref{greatFormula3}) 
we can get a more quantitative handle on the diquark mass
differences.  They begin as equations for differences between the three-halves power of the masses.   From the mass-difference between $N(1680)$ and $\Delta(1950)$ we find
\begin{equation}
(ud)^{3/2} - [ud] ^{3/2} = \frac{2^{1/4}}{\kappa} (1.950-1.680) = .28~ {\rm GeV}^{3/2}
\end{equation}
From this, we see that $(ud)-[ud]$ itself ranges from 360 to 240 MeV as $[ud]$ ranges from 100 to 500 MeV.   
This constitutes a powerful indication of the importance of these diquark correlations, since such energies are quite large in the context of hadron physics.  

A similar comparison among hyperons involves $\Sigma(2030)$ and $\Sigma(1915)$ and leads to 
\begin{equation}
(us)^{3/2} - [us] ^{3/2} = \frac{2^{1/4}}{\kappa} (2.030-1.915) = .12~ {\rm GeV}^{3/2}
\end{equation}
From this, we see that $(us)-[us]$ itself ranges from 150 to 100 MeV as $[us]$ ranges from 200 to 600 MeV.   This is smaller than $(ud)-[ud]$, as expected, but still a very significant energy.  

A more adventurous comparison is to mesons.   Since the same sort of picture, with flux tubes joining weakly coupled ends and feeble spin-orbit forces, works very well 
for them too, 
we are encouraged by the data to compare diquark-quark to
antiquark-quark configurations.  (By the way, this baryon-meson parallelism poses a challenge for Skyrme model or large $N$ approaches to modelling hadrons, since these approaches treat 
mesons and baryons on vastly different footings.)  To be concrete, let us continue to consider orbital angular momentum $L=2$ states with maximal spin and orbital alignment.  
They are as follows: 
\begin{itemize}
\item $[ud]-u: N(1680)$
\item $(ud)-u: \Delta(1950), N(1990)$
\item $[ud]-s: \Lambda(1820)$
\item $[us]-u: \Sigma(1915)$
\item $(us)-u: \Sigma(2030)$
\item $\bar s - u: K_3^{*}(1780)$
\item $\bar d - u: \rho(1690), \omega(1670)$
\item $\bar s -s: \phi(1850)$
\end{itemize}
Now a remarkable thing that appears here, upon comparing the first line with the seventh, or the third with the sixth, is that the mass of the good diquark $[ud]$ is roughly the same as that of $u$ itself!     
This comparison is somewhat
contaminated by tunneling and mixing effects (e.g., tunneling induces mixing between $[ud]-s$ and $[us]-d$), but it's a striking -- and by no means isolated -- phenomenon that 
at large $L$, there is a marked convergence
between mesons and baryons.   Another interesting qualitative pattern is $(ud) > [us] > s > [ud]$.   

The near-equality between effective $[ud]$ and $u$ effective masses, inferred in this way, contrasts with what appears at low $L$, even for heavy quark systems, e.g. 
$\Lambda_{c}(2625)$ versus $D(2460)$ at $L=1$ are split substantially.  On the other hand, this difference of 165 MeV is far less than the conventional ``constituent quark'' mass
$\sim 300$ MeV, and also far less than the 275 MeV difference between $\Lambda_{c}(2285)$ and $D^{*}(2010)$ at $L=0$.  
(Note that heavy-quark hadrons are only half as stretched as their light-quark analogues, for the same $L$, so $L=1$ is ultra-minimal.)    
Part of the reason, I suspect, is that the stretched flux tubes we encounter at larger $L$ can be terminated more smoothly on diquarks, which are extended objects, than on single quarks; this gives the
diquarks an additional energetic advantage.  Another part is simply that the $c$ spin somewhat interferes with the $[ud]$ correlation, and spatial separation lessens this effect.   

Altogether, the concept of diquarks as objects appears to emerge quite naturally and inescapably as an organizing principle for hadron spectroscopy.    As we examine it more carefully, we find that the
energies in play are very significant quantitatively, and that several qualitative refinements with interesting physical interpretations suggest themselves.     
It would be wonderful to illuminate these effects further by numerical experiments in lattice gauge theory.  
The simplest way to see diquark dynamics is just to look at two quarks coupled to a static color source, and in this way to
compare the energy of different spin configurations.  It would be very desirable to verify the strong dependence of the splitting on the quarks' masses.   One could also study the diquark repulsion, by
bringing together the static sources of two such source-diquark systems.   Although it seems very difficult to simulate spinning systems using known techniques of lattice gauge theory, one could study 
quark and diquark systems ``in isolation'' (attached to a flux tube) by artificially introducing a position-dependent mass for the light quarks, that becomes large outside a pocket wherein it vanishes.  This would,
by pushing the quarks away from the source, 
mimic the effect of a centrifugal force.   With insight gained from such studies, 
we would be empowered not only to connect the spectroscopic regularities to foundational QCD, but
also to do better justice to the other fundamental dynamical questions that this circle of ideas wants to encompass.

\bigskip

{\bf Acknowledgment:}  Whatever is new in this paper emerged from joint work with Alex Selem, of which a more complete description is in preparation.   I have also greatly benefitted from 
discussions and earlier collaborative work on related subjects with Robert Jaffe, and from ongoing conversations with Rich Brower and John Negele.  

\bigskip


 \end{document}